\pdfoutput=1

\documentclass{aastex63}

\usepackage[english]{babel} 
\usepackage{hyperref}
\usepackage{upgreek}
\usepackage{amsmath}
\usepackage{amssymb}
\usepackage[version=3]{mhchem}

\usepackage{graphicx}
\usepackage{epstopdf}
\graphicspath{{figures/}} 
\usepackage{tabu}
\usepackage{soul}
\usepackage{color}

\newcommand{\NaMg}{$^{23}\mathrm{Na}(\alpha,p)^{26}\mathrm{Mg}$ }

\begin{document}
\NewPageAfterKeywords

\title{New experimental \texorpdfstring{\NaMg{}}{23Na(a,p)26Mg} Reaction Rate for Massive Star and Type-Ia Supernova  models}

\accepted{11-Mar-2021}
\submitjournal{ApJ}

\author{N. J. Hubbard}
\affiliation{Department of Physics, University of York, York YO10 5DD, UK}
\affiliation{Department of Physics and Astronomy, Aarhus University, 8000 Aarhus C, Denmark}
\affiliation{Technische Universit\"{a}t Darmstadt, Insitit f\"{u}r Kernphysik, 64289 Darmstadt, Germany}
\affiliation{GSI Helmholtzzentrum f\"{u}r Schwerionenforschung, 64291 Darmstadt, Germany}

\author{C. Aa. Diget}
\affiliation{Department of Physics, University of York, York YO10 5DD, UK}

\author{S. P. Fox}
\affiliation{Department of Physics, University of York, York YO10 5DD, UK}

\author{H. O. U. Fynbo}
\affiliation{Department of Physics and Astronomy, Aarhus University, 8000 Aarhus C, Denmark}

\author{A. M. Howard}
\affiliation{Heinz Maier-Leibnitz Zentrum (MLZ),  FRM-II,  Technische Universit\"{a}t M\"{u}nchen, 85748 Garching, Germany}

\author{O. S. Kirsebom}
\affiliation{Institute for Big Data Analytics, Dalhousie University, Halifax, Nova Scotia B3H 4R2, Canada}

\author{A. M. Laird}
\affiliation{Department of Physics, University of York, York YO10 5DD, UK}

\author{M. Munch}
\affiliation{Department of Physics and Astronomy, Aarhus University, 8000 Aarhus C, Denmark}

\author{A. Parikh}
\affiliation{Department de F\'{i}sica, Universitat Polit\`{e}cnica de Catalunya, E-08036 Barcelona, Spain}

\author{M. Pignatari}
\affiliation{E. A. Milne Centre for Astrophysics, University of Hull, Hull, HU6 7RX, UK}
\affiliation{Konkoly Observatory, Research Centre for Astronomy and Earth Sciences, Hungarian Academy of Sciences, Konkoly Thege Miklos ut 15-17, H-1121 Budapest, Hungary}
\affiliation{Joint Institution for Nuclear Astrophysics -- Center for the Evolution of the Elements, USA}
\affiliation{NuGrid Collaboration \url{https://nugrid.github.io}}

\author{J. R. Tomlinson}
\affiliation{Department of Physics, University of York, York YO10 5DD, UK}

\nocollaboration{50}

\correspondingauthor{N. J. Hubbard}
\email{n.hubbard@gsi.de}


\begin{abstract}
	The \NaMg{} reaction has been identified as having a significant impact on the nucleosynthesis of several nuclei between Ne and Ti 
	in type-Ia supernovae, 
	and of $^{23}$Na and $^{26}$Al in massive stars. The reaction has been subjected to renewed experimental interest recently, motivated by high 
	uncertainties in early experimental data and in the statistical Hauser-Feshbach models used in 
	reaction rate compilations. 	
	Early experiments were affected by target deterioration issues and unquantifiable uncertainties. Three new independent measurements instead are utilizing inverse kinematics and Rutherford scattering monitoring to resolve this.
	In this work we present directly measured angular distributions of the emitted protons
	to eliminate a discrepancy in the assumptions made in the recent reaction rate measurements, which results in cross sections differing by a factor of 3. We derive a new combined experimental reaction rate for the \NaMg reaction with a total uncertainty of 30\% at relevant temperatures.
	Using our new \NaMg rate, the $^{26}$Al and $^{23}$Na production uncertainty is reduced to within 8\%. In comparison, using the factor of 10 uncertainty previously recommended by the rate compilation STARLIB, $^{26}$Al and $^{23}$Na production was changing by more than a factor of 2. In type-Ia supernova conditions, the impact on production of $^{23}$Na is constrained to within 15\%.
\end{abstract}


\section{Introduction}

The \NaMg{} reaction rate has been the subject of a number of nuclear astrophysics studies in the last decade. During carbon fusion in massive stars, $^{23}$Na is produced directly by one of the main fusion channels, $^{12}$C$(^{12}$C$,p)^{23}$Na. Sodium made in this phase will provide the bulk of the sodium ejected by core-collapse supernovae, which are an important source of sodium in the Universe \citep[e.g.,][]{Timmes1995, woosley:02,Kobayashi2011,Pignatari2016,sukhbold:16,Limongi2003,Chieffi2013}. In these conditions, the $^{23}$Na(p,$\alpha$)$^{20}$Ne is the main destruction channel for $^{23}$Na. However, among other reactions \NaMg{} is also fully activated and needs to be considered to calculate stellar abundances.  
For instance, \cite{Iliadis2011} showed that a factor of 10 increase in the \NaMg{} reaction rate would increase the $^{26}$Al production by a factor of 3, acting as a source of protons for the $^{25}$Mg$(p,\gamma)^{26}$Al reaction.
$^{26}$Al is a key radioisotope for astronomy. Its decay signature has been measured in the interstellar-medium by $\gamma$-ray telescopes COMPTEL \citep{Chen1995} and INTEGRAL \citep{Diehl2012}, and its abundance derived from meteoritic material is used as a fundamental diagnostic to study the formation of the solar system \citep[][and references therein]{lugaro:18}.
The reaction has also been identified as affecting the production of $^{21}$Ne, $^{23}$Na, $^{26}$Mg, $^{29}$Si, $^{43}$Ca and $^{47}$Ti, in type-Ia supernovae \citep{BravoMartinez2012}, with most of the abundances changing by a factor of 0.12 -- 2 when the \NaMg{} reaction rate is increased or decreased by a factor of 10, and with $^{26}$Mg and $^{43}$Ca changing by over a factor of 2. The impact on  $^{23}$Na has also been identified in the deflagration-detonation transition supernova (DDT-SN) model of \cite{Parikh2013}, where an increase of the \NaMg{} rate by a factor of 10 yields a change in abundance by a factor of 0.47.

The reaction has been subjected to renewed experimental interest recently, due to identified weaknesses in early experimental data and uncertainties in the application of statistical Hauser-Feshbach models \citep{Iliadis2011,PARIKH2013225}.
There now exist three new independent measurements of the reaction cross-section utilizing novel techniques to solve the issues of early experiments. Two of these experiments \citep{AlmarazCalderon2015,Tomlinson2015} are only capable of detecting a small angular range of the emitted protons, necessitating an assumption on the variation of the reaction's cross-section with respect to angle (angular distribution) in order to calculate a total cross-section for astrophysical purposes. The two measurements assumed different distributions. The third measurement \citep{Howard2015} has sufficient angular coverage to directly measure the angular distribution of the protons.
In the present paper we show that when
corrected for the angular distributions presented, these measurements are all consistent to within 30\% with one another in the energy range $E_{cm} = 1.7 - 3.0$ MeV.

The angular distributions have been obtained from the data taken at Aarhus University \citep{Howard2015}, which can be applied to the other data sets to eliminate angular distribution assumptions, reducing systematic uncertainties on their cross-section measurement, and ensuring consistency across the three measurements.
These new angular-distribution-corrected measurements have been combined to obtain a new recommended experimental astrophysical reaction rate, with a significantly reduced uncertainty. The impact of this rate has been modeled in massive stars and type-Ia supernova.

\section{Experimental History}

\begin{figure}
	\plotone{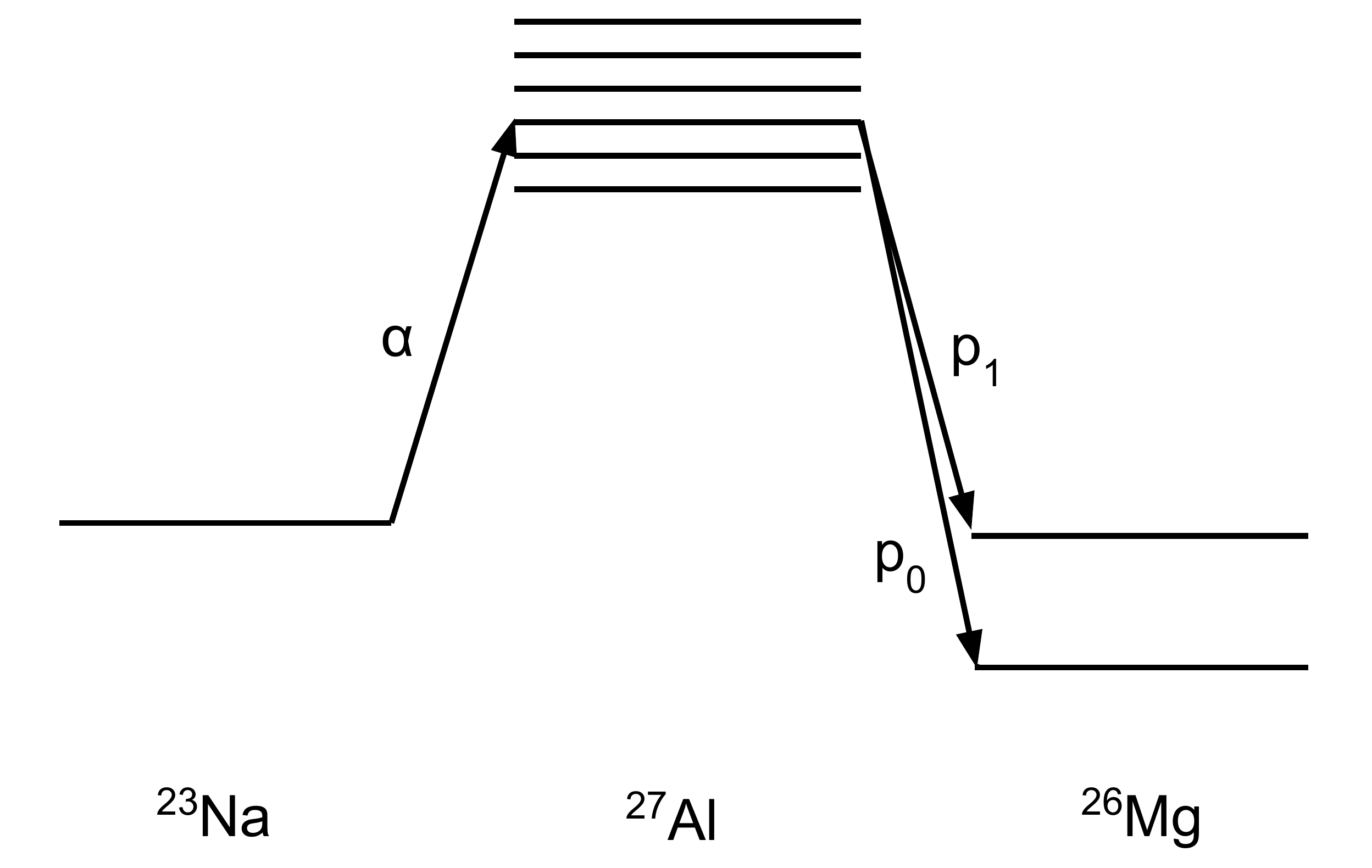}
	\caption{\label{fig:ReactionDiagram}
		Schematic of the \NaMg reaction. Alpha particles react with $^{23}$Na to form $^{27}$Al in an excited state, which then decays to $^{26}$Mg emitting a proton. If $^{26}$Mg is produced in the ground state, the corresponding protons are called $p_0$ protons. For $^{26}$Mg in its first excited state, they are $p_1$ protons.
	}
\end{figure}

A schematic of the \NaMg reaction is shown in figure \ref{fig:ReactionDiagram}: a $^{23}$Na ion reacts with an $\alpha$ particle ($^{4}$He) producing a compound $^{27}$Al nucleus in an excited state. This excited compound nucleus rapidly decays into $^{26}$Mg by emitting a proton. If $^{26}$Mg is produced in its ground state the protons are called $p_0$ protons, and if in its first excited state $p_1$ protons.

The earliest measurement of the \NaMg reaction was by \cite{Kuperus1964} who measured individual resonance strengths ($\omega\gamma$)
of narrow resonances between $E_\alpha = 1-3.3$ MeV, producing $^{26}$Mg in its ground state ($p_0$ protons).
This was followed up by \cite{Whitmire1974} who measured resonance strengths producing $^{26}$Mg in its first excited state ($p_1$ protons) as well as the ground state. They additionally calculated a reaction rate based on their data.
Because Whitmire \& Davids did not reassess the $p_0$ resonance strengths at the energies already published by Kuperus, the combined work which incorporates both $p_0$ and $p_1$ cross-sections (derived from the published resonance strengths) will be treated as one data set in the subsequent discussion. Both experiments were in forward kinematics, impinging a $^4$He beam on a NaCl target.

The experiments involved observation of individual resonances, and therefore the targets used were thin ($\approx 25\;\mu$g/cm$^2$) and measurements were taken with small increases in beam energy. Thus, the beam current of the incoming $\alpha$ particles was fairly high ($\approx 400$ nA) in order to collect sufficient statistics. This unfortunately led to degradation of the targets due to heating of the target and the low melting point of NaCl of $801 ^\circ$C.  This effect was discussed by \cite{Whitmire1974}, but was not quantified, leading to a large systematic uncertainty in the data. Subsequently reaction rate compilations such as REACLIB \citep{JINAREACLIB} therefore used cross-sections and reaction rates predicted from statistical Hauser-Feshbach (HF) models.

A detailed sensitivity study of $^{26}$Al production was performed in 2011 by \cite{Iliadis2011},
which identified the \NaMg reaction as having a significant effect on $^{26}$Al production in massive star models
(a factor of 10 in the rate affects $^{26}$Al production by a factor of 3).
The experimental limitations and reliance on statistical models were also highlighted, including limitations on the applicability of HF to low-mass, alpha-induced, reactions such as the \NaMg rate \citep{Mohr2015}.
A comparison of HF to experimental reaction rates by \cite{PARIKH2013225} noted a large number of values which exceed the traditional factor of 2 reliability of HF. Based on the study, further experimental work was strongly advised. 

The first modern measurement on the \NaMg reaction was performed by \cite{AlmarazCalderon2014,AlmarazCalderon2015} at Argonne National Laboratory (ANL) near Chicago, USA. This measurement was performed in inverse kinematics, utilizing a $^{23}$Na beam and a cryogenic $^{4}$He gas target. The use of inverse kinematics avoided the problems of target
deterioration observed by Kuperus and Whitmire \& Davids.
Due to the nature of the experimental setup only protons emitted at angles of $\theta_{cm} \ge 160^\circ$ were measured.
In order to obtain a full cross-section, the angular distribution of the protons
was assumed to be the same as the $^{27}$Al(a,p)$^{30}$Si reaction. The initial results indicated cross-sections much higher than expected from Kuperus and Whitmire \& Davids or HF models, but these data were subsequently re-analyzed producing cross-sections consistent with the HF cross-sections. 

Three additional experiments on the \NaMg reaction were performed soon after the first result, the first, by \cite{Howard2015} at Aarhus University in Denmark, in forward kinematics.
By utilizing a much lower beam intensity ($200-500$ ppA) than
Kuperus and Whitmire \& Davids while also monitoring Rutherford scattered alpha particles to monitor target
deterioration, the large source of uncertainty from the target stoichiometry was eliminated.
This experiment was also able to directly measure the proton angular distributions, which are presented in the next section.

The second experiment, by \cite{Tomlinson2015}, was performed in inverse kinematics at TRIUMF in Canada. This experiment had a very similar
experimental set-up to that of \cite{AlmarazCalderon2014}, but with a room-temperature target and a broader angular coverage and assuming an isotropic angular distribution.
Both \cite{Howard2015} and \cite{Tomlinson2015} measured cross-sections consistent with those of the NON-SMOKER HF model \citep{Rauscher2000}.

The third experiment was performed by \cite{Aliva2016}, also at ANL. This measurement utilised an active target measurement to directly measure the total cross-section of the \NaMg reaction, in inverse kinematics, covering an energy range of $2 -6$ MeV in the centre-of-mass frame. The cross-sections of this experiment were also consistent with those of NON-SMOKER.

\begin{figure}
	\includegraphics[width=0.95\linewidth]{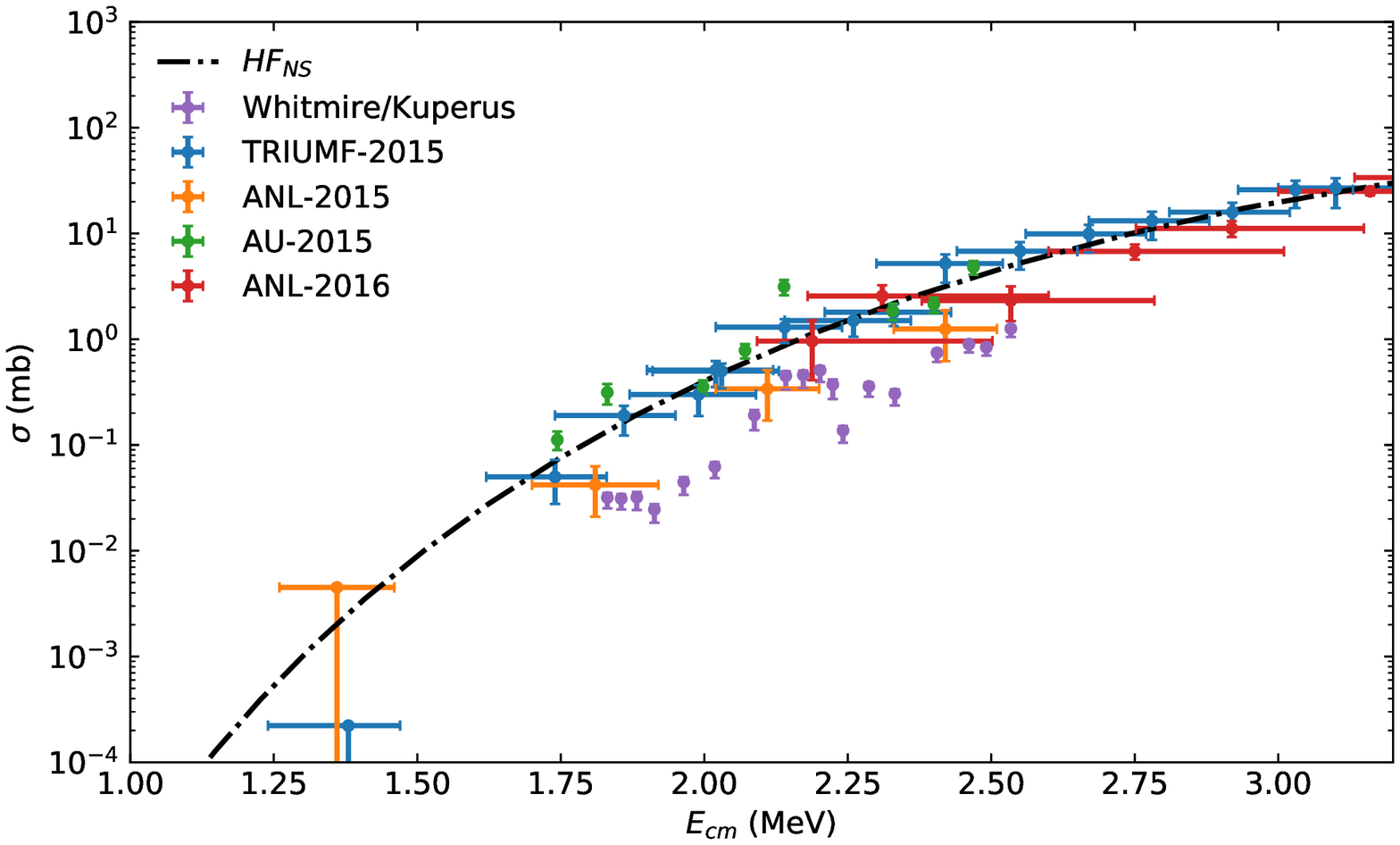}
	\caption{\label{fig:PublishedCrossSections}
	Published cross sections for the \NaMg reaction from \cite{Tomlinson2015} (TRIUMF-2015), \cite{AlmarazCalderon2015} (ANL-2015), \cite{Howard2015} (AU-2015) and \cite{Avila2016} (ANL-2016), and Hauser-Feshbach calculated cross-sections ($HF_{NS}$) \citep{Rauscher2000}.
	}
\end{figure}

The published cross-sections for the five data sets are shown in figure \ref{fig:PublishedCrossSections}. All four recent measurements of the \NaMg reaction
have reasonably consistent results, with \cite{Tomlinson2015} and \cite{Howard2015} agreeing within 50\%. The data from \cite{AlmarazCalderon2015} can be seen to be systematically lower by a factor 2.5 from the other two data sets. The newly analyzed angular distribution data from AU-2015 can be applied to TRIUMF-2015 and ANL-2015 to eliminate the discrepancy between the three data sets, as detailed in the following.

\section{Experimental Angular Distributions}

The angular distributions of the emitted protons produced by the \NaMg reaction were measured at Aarhus University, Denmark as part of the direct measurement of the cross-sections, and the experimental set-up is described in detail by \cite{Howard2015}. The detector array comprises of two detectors covering the laboratory angular range of $60^\circ-120^\circ, 140^\circ-164^\circ$. Protons are unambiguously identified at $E_{cm} > 1.9$ MeV, while $p_1$ protons below this energy were only measured at $140^\circ-164^\circ$.

For the higher-energy measurements a full angular distribution was obtained by fitting a sum of the first four even Legendre polynomials: 

\begin{equation}
\frac{\mathrm{d}\sigma}{\mathrm{d}\Omega} = a + b P_2(\cos(\theta)) + c P_4(\cos(\theta)) + d P_6(\cos(\theta))
\end{equation}

which can then be integrated over all angles to obtain a total cross-section. These distributions are symmetric around $90^\circ$, which is expected for a reaction involving an intermediate compound nucleus, but a conservative uncertainty of 20\% was introduced nevertheless.
For the lowest energies where the $p_1$ protons could only be measured at high angles, an isotropic distribution was assumed. This adds a systematic uncertainty of 30\% to the cross-sections at $E_{cm} < 1.9$ MeV. A full account of the uncertainty budget is detailed by \cite{Howard2015}.

\begin{figure}
	\includegraphics[width=0.95\linewidth]{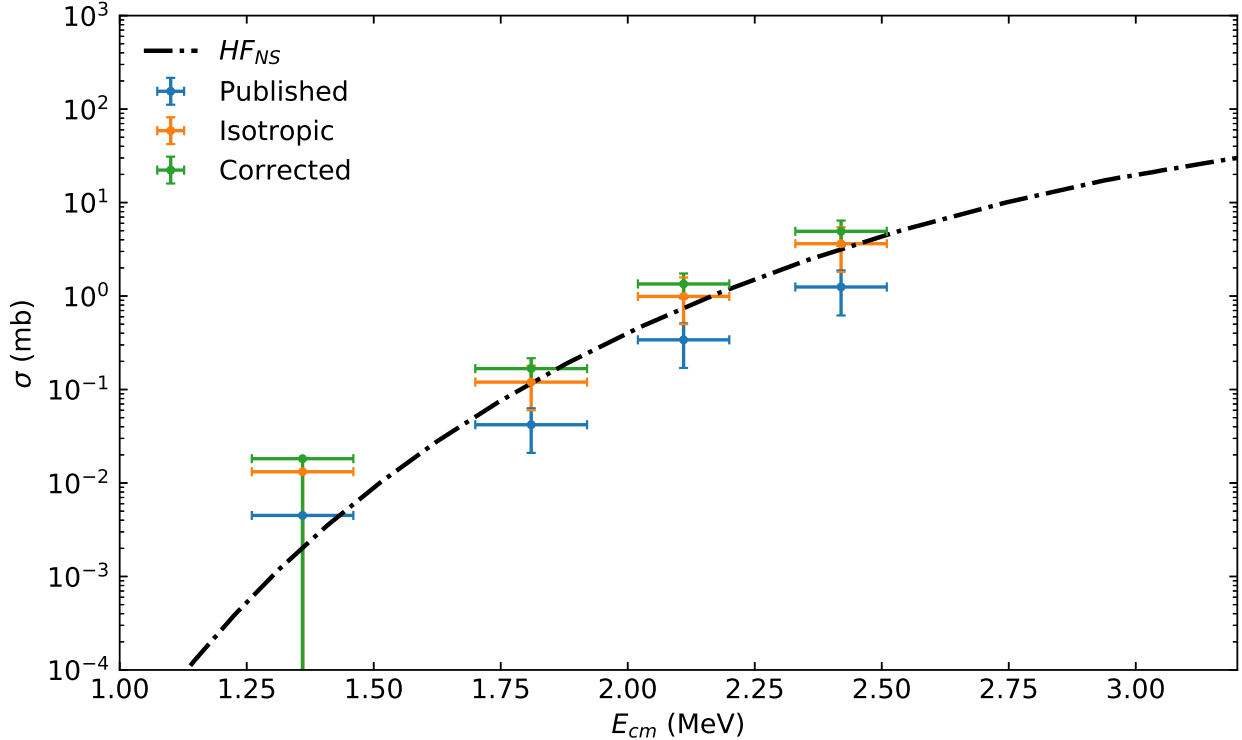}
	\caption{
	\label{fig:AnlCrossSections}
	Comparisons of the cross-sections from \cite{AlmarazCalderon2015} (ANL-2015) and after correcting to an isotropic distribution and the experimentally measured angular distribution calculated in this work. 
	}	
\end{figure}

\begin{figure}
	\includegraphics[width=0.95\linewidth]{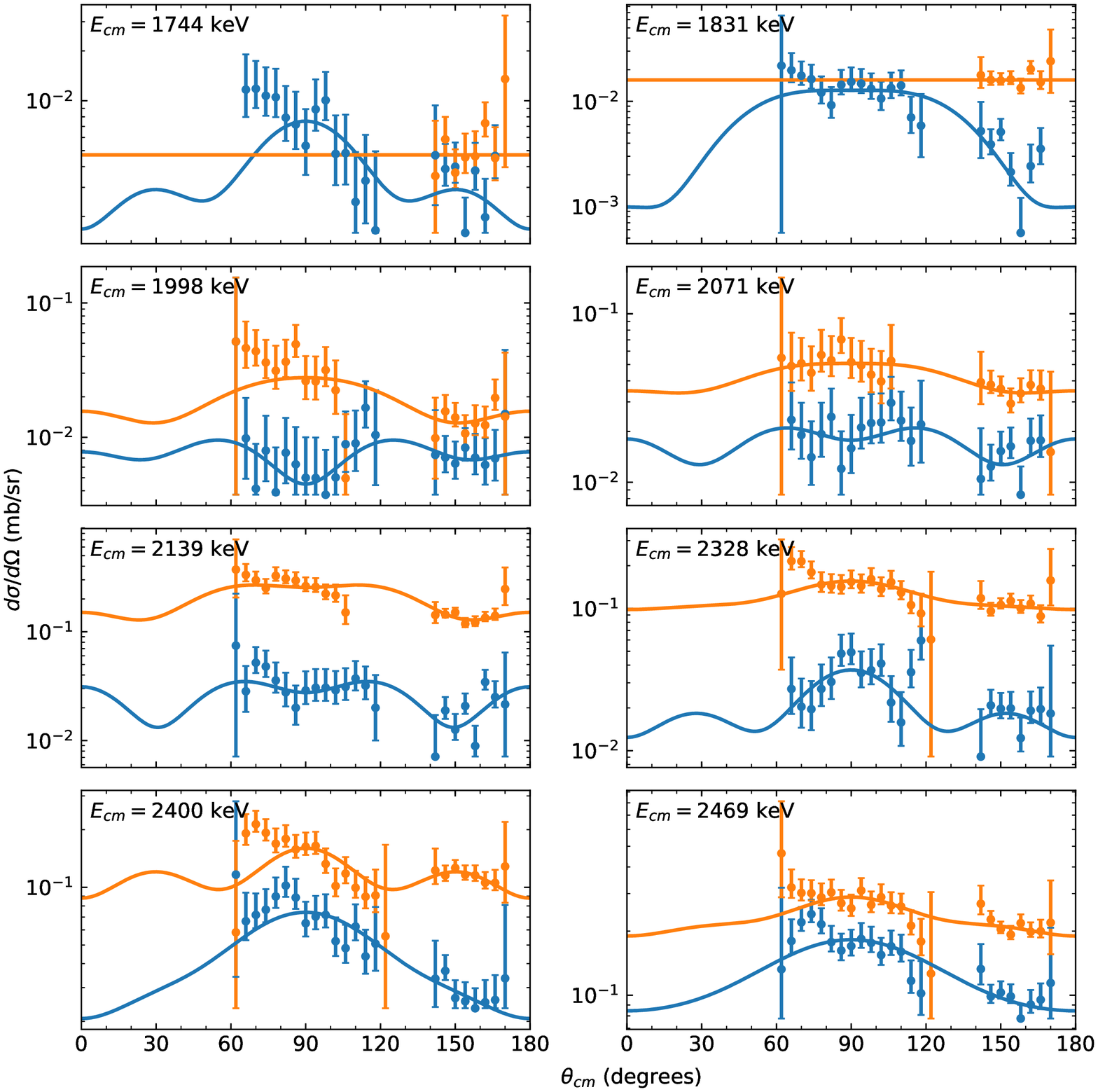}
	\caption{Directly measured angular distributions for the $p_1$ (orange/grey) and $p_0$ (blue/black) protons. The differential cross-sections measured at Aarhus University are fitted with a sum of the first four even Legendre polynomials (solid line) which can then be integrated over all angles to obtain a total cross-section. The  energies $E_{cm} = 1744$ and $1831$ keV did not have data below $140^\circ$ for $p_1$ protons, thus an isotropic distribution was used instead.
	\label{fig:AllDistributions}}
\end{figure}

All energies were additionally fitted to isotropic distributions using only the high-angle data. This allows evaluation of the systematic uncertainty on the low-energy points discussed above, and application of the distributions from the Howard data to the TRIUMF-2015 and ANL-2015 data, by applying the ratio of measured to isotropic cross-sections to their isotropic cross-sections. For ANL-2015 the total cross-sections were calculated from their differential cross-sections, rather than applying the angular distribution assumed in the paper \citep{AlmarazCalderon2015}, a comparison of the effect of the different angular distributions for ANL-2015 is shown in figure \ref{fig:AnlCrossSections}. All evaluated cross-sections using the measured distributions ($\sigma$) and the cross-sections assuming an isotropic distribution ($\sigma^{ISO}$) are shown in table \ref{tbl:IntegratedCrossSections}, where it can be seen that on average the cross-sections are 30\% higher than under an isotropic approximation. Although the angular distribution impact should depend on the incident beam energy, the thick targets employed by all the measurements averages out these effects. This can be noted by the cross-section impact having no significant energy dependence, with the exception of the point at $E_{cm} = 1831$ keV. This factor of 4.88 increase at $E_{cm} = 1831$ keV is due to a strong resonance at $1800$ keV producing a nearly pure $\ell = 1$ distribution which drops off strongly at angles above $120^\circ$. The angular distributions are shown in figure \ref{fig:AllDistributions}. For the subsequent analysis an energy-independent increase of 30\% over the isotropic assumption is used to correct the data from ANL-2015 and TRIUMF-2015.

\newcommand{\asyme}[3]{$#1(^{+#2}_{-#3})$}

\begin{deluxetable*}{ccccccc}
	
	\tablecaption{Angle-integrated cross-sections and their total errors, compared to cross-sections assuming an isotropic angular-distribution $(\sigma^{ISO})$. The error budget is discussed in detail elsewhere \citep{Howard2015}\label{tbl:IntegratedCrossSections}}

	\tablehead{
	\colhead{$E_{cm}$ (keV)} &
	\colhead{$\sigma_{p0}$ (mb)} &
	\colhead{$\sigma_{p0}^{ISO}$ (mb)} &
	\colhead{$\sigma_{p0}/\sigma_{p0}^{ISO}$} &
	\colhead{$\sigma_{p1}$ (mb)} &
	\colhead{$\sigma_{p1}^{ISO}$ (mb)} &
	\colhead{$\sigma_{p1}/\sigma_{p1}^{ISO}$}
	}

	\startdata
	1744 & 0.05(1)  & 0.04 & 1.5(4) & \asyme{0.06}{2}{2}   & --- & --- \\
	1831 & 0.11(2)  & 0.02 & 4.9(14) & \asyme{0.20}{6}{6}   & --- & --- \\
	1998 & 0.09(2)  & 0.09 & 1.1(2) & \asyme{0.26}{5}{6}   & 0.17 & 1.6(3) \\
	2071 & 0.22(4)  & 0.17 & 1.3(3) & \asyme{0.56}{11}{12} & 0.43 & 1.3(3) \\
	2139 & 0.34(7)  & 0.20 & 1.7(4) & \asyme{2.80}{56}{62} & 1.69 & 1.7(4) \\
	2328 & 0.28(6)  & 0.22 & 1.2(3) & \asyme{1.58}{32}{35} & 1.30 & 1.2(3) \\
	2400 & 0.62(12) & 0.34 & 1.8(4) & \asyme{1.52}{30}{33} & 1.45 & 1.0(2) \\
	2469 & 1.76(35) & 1.19 & 1.5(3) & \asyme{3.05}{61}{67} & 2.61 & 1.2(2) \\
	\enddata
	
\end{deluxetable*}

\section{Experimental Reaction Rate}

With the three recent cross-section measurements we obtain a new experimental reaction rate which can be used in preference to the Hauser-Feshbach (HF) reaction rate, removing the uncertainties associated with the models. Before combining the three measurements, we apply the angular distributions discussed above to \cite{AlmarazCalderon2015} and \cite{Tomlinson2015}, reducing uncertainties associated with the assumptions of angular distributions. These corrections result in a $30\%$ increase over the isotropic distribution assumed by \cite{Tomlinson2015}, and a $250\%$ increase 

\begin{figure}
	\includegraphics[width=0.95\linewidth]{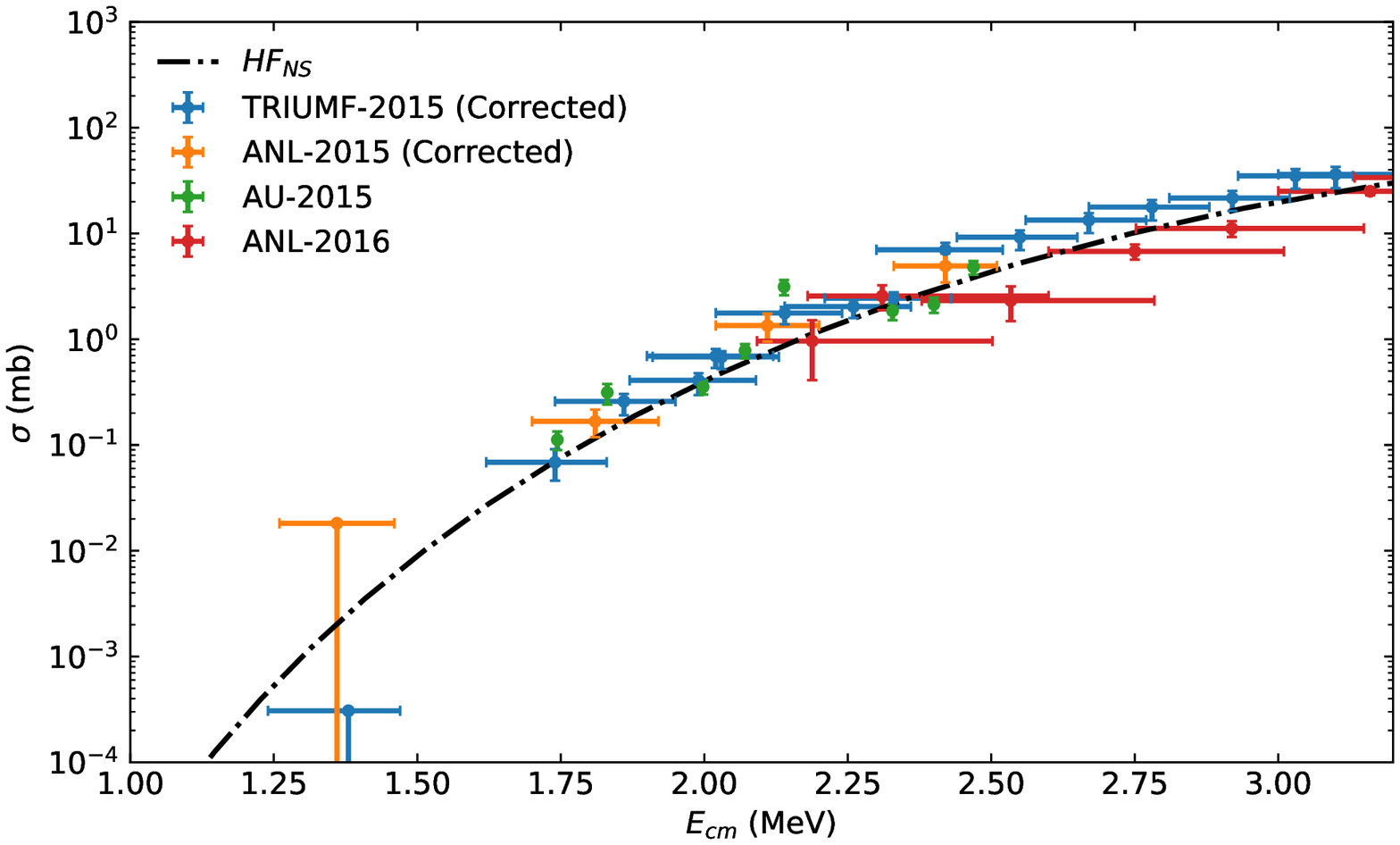}
	\caption{
	\label{fig:CorrectedCrossSections}
	Cross-sections from \cite{Tomlinson2015} (TRIUMF-2015), 
	\cite{AlmarazCalderon2015} (ANL-2015), \cite{Howard2015} (AU-2015) and and \cite{Avila2016} (ANL-2016) with the TRIUMF-2015 and ANL-2015 cross-sections corrected for the angular distributions directly measured at AU-2015 as discussed in the text. These corrections increase the $p_0$ cross-sections by a factor of 1.4(3) and the $p_1 $ cross-sections by a factor of 1.3(2) over the isotropic distribution assumed in TRIUMF-2015 and increase in the $p_0$ and $p_1$ cross-sections of factors of 4.3(9) and 3.8(7) over the $^{27}$Al$(\alpha,p)^{30}$Si distribution assumed in ANL-2015.
	}	
\end{figure}

\begin{figure}
	\includegraphics[width=0.95\linewidth]{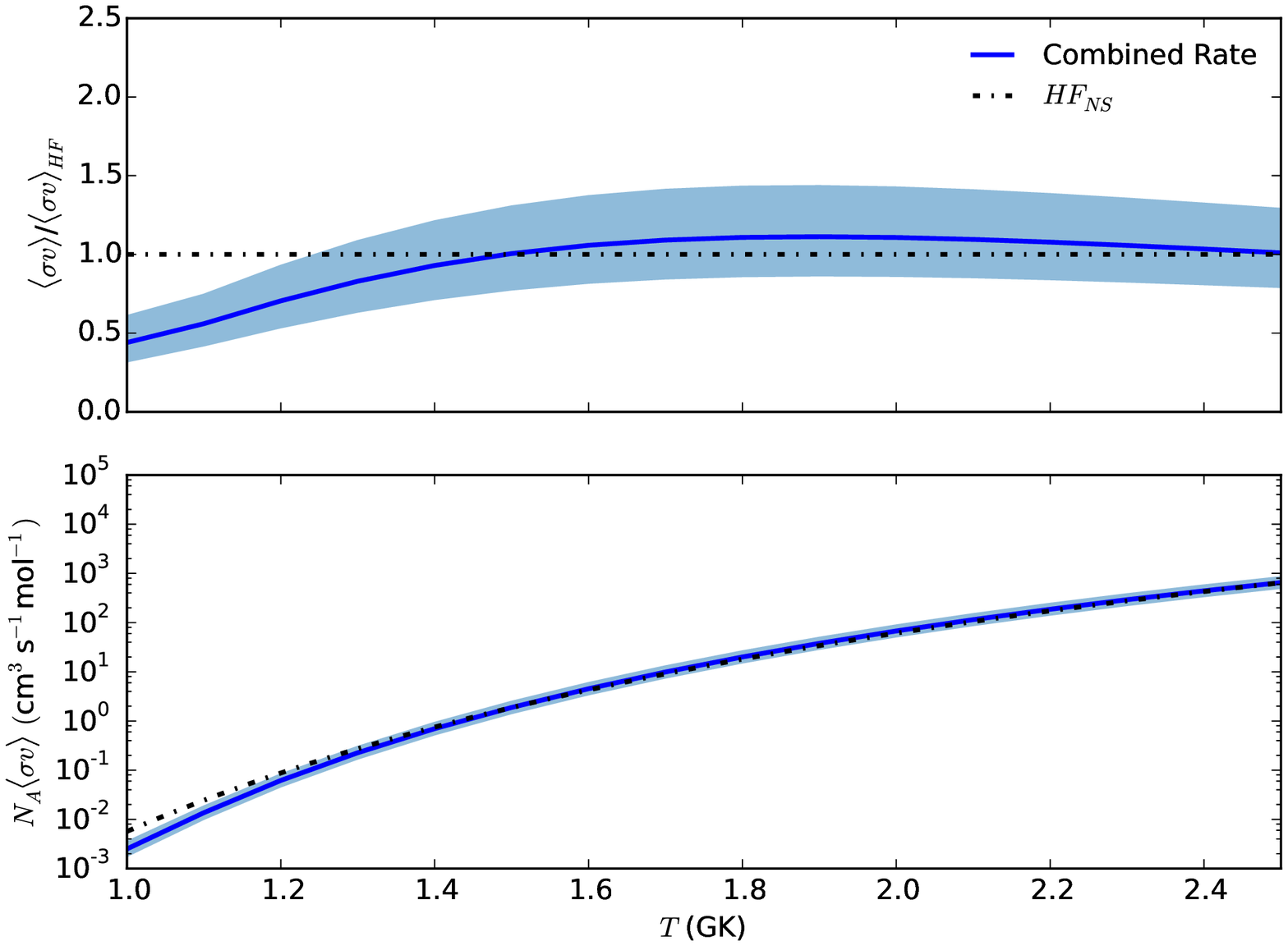}
	\caption{Experimental reaction rate for the \NaMg reaction
		derived from averaging reaction rates calculated using \textsc{exp2rate} for the three corrected experimental cross-sections. The top figure compares the new rate to the previous recommended rate from NON-SMOKER calculations, where it agrees within uncertainty at higher temperatures, but drops below the HF rate below 1.2 GK. The bottom figure shows the absolute reaction rate and its temperature dependence. 
		\label{fig:CombinedReactionRate}}
\end{figure}

Each of the cross-sections was then integrated to obtain a reaction rate using \textsc{exp2rate} \citep{exp2rate}. This code takes as
input center-of-mass energies ($E_{cm}$), their cross sections ($\sigma)$,
and the associated cross-section uncertainties. It then converts these cross sections into
astrophysical $S$ factors before numerically integrating to obtain the reaction rate with errors. The reaction rate is given tabulated over a range of temperatures.
These three individual tabulated rates are then combined by taking a weighted average of the three. Combining the rates in this way avoids issues with the different cross-section energies measured in the three experiments. For the data point at $E_{cm} = 1.38$ MeV, only the more constrained limit by TRIUMF-2015 is taken into account, being used for the upper limit and recommended reaction rate, and a cross-section of 0 at this energy is used for the lower limit of the combined reaction rate.
The new experimental reaction rate and its uncertainty is shown in figure \ref{fig:CombinedReactionRate}, and tabulated in table \ref{tbl:Rate} The rate agrees within uncertainty with the HF predicted rate, but deviates to below the HF rate as temperatures under 1.2 GK, due to the upper-limit of the TRIUMF-2015 data point at $E_{cm} = 1.38$ MeV.

\begin{deluxetable*}{cccc}
	\tablecaption{Experimental \NaMg reaction rate in units of cm$^3$ s$^{-1}$ mol$^{-1}$
		\label{tbl:Rate}}
	\tablehead{
		\colhead{$T$ (GK)} &
		\colhead{Recommended Rate} &
		\colhead{Lower Limit} &
		\colhead{Upper Limit}
	}
	
	\startdata
	$1.0$ &  $2.5\times10^{-3}$ &  $1.8\times10^{-3}$ &  $3.5\times10^{-3}$ \\
	$1.1$ &  $1.4\times10^{-2}$ &  $1.0\times10^{-2}$ &  $1.8\times10^{-2}$ \\
	$1.2$ &  $6.2\times10^{-2}$ &  $4.7\times10^{-2}$ &  $8.1\times10^{-1}$ \\
	$1.3$ &  $2.3\times10^{-1}$ &  $1.7\times10^{-1}$ &  $3.0\times10^{-1}$ \\
	$1.4$ &  $7.0\times10^{-1}$ &  $5.4\times10^{-1}$ &  $9.1\times10^{-1}$ \\
	$1.5$ &  $1.9\times10^{ 0}$ &  $1.5\times10^{ 0}$ &  $2.4\times10^{ 0}$ \\
	$1.6$ &  $4.5\times10^{ 0}$ &  $3.5\times10^{ 0}$ &  $5.9\times10^{ 0}$ \\
	$1.7$ &  $9.9\times10^{ 1}$ &  $7.7\times10^{ 0}$ &  $1.3\times10^{ 1}$ \\
	$1.8$ &  $2.0\times10^{ 1}$ &  $1.6\times10^{ 1}$ &  $2.6\times10^{ 1}$ \\
	$1.9$ &  $3.8\times10^{ 1}$ &  $3.0\times10^{ 1}$ &  $4.9\times10^{ 1}$ \\\
	$2.0$ &  $6.8\times10^{ 1}$ &  $5.3\times10^{ 1}$ &  $8.7\times10^{ 1}$ \\
	$2.1$ &  $1.1\times10^{ 2}$ &  $9.0\times10^{ 1}$ &  $1.5\times10^{ 2}$ \\
	$2.2$ &  $1.9\times10^{ 2}$ &  $1.5\times10^{ 2}$ &  $2.4\times10^{ 2}$ \\
	$2.3$ &  $2.9\times10^{ 2}$ &  $2.3\times10^{ 2}$ &  $3.7\times10^{ 2}$ \\
	$2.4$ &  $4.4\times10^{ 2}$ &  $3.5\times10^{ 2}$ &  $5.7\times10^{ 2}$ \\
	$2.5$ &  $6.5\times10^{ 2}$ &  $5.1\times10^{ 2}$ &  $8.3\times10^{ 2}$ \\
	$2.6$ &  $9.3\times10^{ 2}$ &  $7.3\times10^{ 2}$ &  $1.2\times10^{ 3}$ \\
	$2.7$ &  $1.3\times10^{ 3}$ &  $1.0\times10^{ 3}$ &  $1.7\times10^{ 3}$ \\
	\enddata	
\end{deluxetable*}

For most astrophysical models, a large reaction database such as REACLIB
\citep{JINAREACLIB} is used. In these databases the reaction rates
are stored as coefficients to the function:
\begin{equation}
\label{eqn:ReaclibCoeffs}
\lambda = \exp{\left[a_0 + \sum_{i=1}^{5} a_i T_9^{\frac{2i-5}{3}} + a_6\ln{(T_9)}\right]},
\end{equation}
which accurately captures the temperature dependence of most reaction rates.
The combined tabulated reaction rates were fitted using the Computation Infrastructure for Nuclear Astrophysics (CINA) system \citep{CINA}, and the coefficients for equation \ref{eqn:ReaclibCoeffs} are tabulated in table \ref{tbl:NaMgReactionCoeffects}

\begin{deluxetable*}{cccc}
	\tablecaption{REACLIB \citep{JINAREACLIB} coefficients of equation \ref{eqn:ReaclibCoeffs} for the \NaMg reaction
		\label{tbl:NaMgReactionCoeffects}}
		\tablehead{
		&
		\colhead{Recommended Rate} &
		\colhead{Lower Limit} &
		\colhead{Upper Limit}
		}
		
		\startdata
		$a_0$ &  $0.789003\times10^{3}$ &  $0.682596\times10^{3}$ &  $0.102587\times10^{4}$ \\
		$a_1$ & $-0.148922\times10^{2}$ & $-0.112997\times10^{2}$ &  $-0.217681\times10^{2}$ \\
		$a_2$ &  $0.104863\times10^{4}$ &  $0.785346\times10^{3}$ & $-0.156395\times10^{4}$ \\
		$a_3$ & $-0.194956\times10^{4}$ & $-0.156219\times10^{4}$ &  -$0.274012\times10^{4}$ \\
		$a_4$ &  $0.130023\times10^{3}$ &  $0.106911\times10^{3}$ & $0.178909\times10^{3}$ \\
		$a_5$ & $-0.920273\times10^{1}$ & $-0.767373\times10^{1}$ &  -$0.125179\times10^{2}$ \\
		$a_6$ &  $0.888444\times10^{3}$ &  $0.696058\times10^{3}$ & $0.127309\times10^{4}$ \\
		\enddata	
\end{deluxetable*}

\begin{deluxetable*}{cccc}
	\tablecaption{REACLIB \citep{JINAREACLIB} coefficients of equation \ref{eqn:ReaclibCoeffs} for the $^{26}$Mg$(p,\alpha)^{23}$Na reverse reaction. Untabulated coefficients are identical to those in table \ref{tbl:NaMgReactionCoeffects}.
		\label{tbl:NaMgRevReactionCoeffects}}
	\tablehead{
		&
		\colhead{Recommended Rate} &
		\colhead{Lower Limit} &
		\colhead{Upper Limit}
	}
	
	\startdata
	$a_0^{rev}$ &  $0.791581\times10^{3}$ &  $0.685174\times10^{3}$ &  $0.102845\times10^{4}$ \\
	$a_1^{rev}$ & $-0.360201\times10^{2}$ & $-0.324276\times10^{2}$ &  $-0.428960\times10^{2}$ \\
	\enddata	
\end{deluxetable*}

The reaction rate for the reverse reaction $^{26}$Mg$(p,\alpha)^{23}$Na can also
be related to the \NaMg reaction rate, and thus can also be computed from the
experimental data and the partition functions ($G$), which are tabulated by \cite{Rauscher2000}.
This reverse rate was also fitted to equation \ref{eqn:ReaclibCoeffs} using
CINA. The coefficients are tabulated in \ref{tbl:NaMgRevReactionCoeffects}, with the unspecified coefficients unchanged from \ref{tbl:NaMgReactionCoeffects}.

\section{Astrophysical Impact}

The impact of this new experimental rate in the C-burning convective shell is
investigated using the \textsc{NuGrid} nucleosynthesis post-processing toolkit \citep{NuGridPaper}.
An extended network of reaction rates is used at given stellar conditions, following consistently the production and destruction of isotopic abundances over time. For a detailed description of the nuclear reaction network and reaction rates used, we refer to \cite{Pignatari2016}. \cite{Iliadis2011} performed the sensitivity study of the \NaMg rate in two massive star models of initial mass $25M_\odot$ and $60M_\odot$, respectively. Therefore, we selected two massive-star models of the same mass from the \cite{Pignatari2016} stellar set. 
The stellar structure is produced by \textsc{GENEC} \citep{Eggenberger2008}, with initial metallicity $Z = 0.02$. 
For this work, we used
the software \textsc{ppn} 
to investigate the variation of isotopic abundances in single trajectories extracted from C-burning regions of the star.

The trajectories
($T$, $\rho$  as a function of time) and initial abundances were taken from the deepest area of the C-burning convective shell. For the $60M_\odot$ star, we extracted the trajectory at mass coordinate $3.5 M_\odot$, between 1.3 and 0.12 years before collapse.

The temperature ranges from $1.1-1.3$ GK and the density $6.0\times10^4-8.6\times10^4$ g cm$^{-3}$.
With this trajectory we used a network of 1100 isotopes
with changes made solely to the \NaMg reaction and its inverse, the rest of the network setup was identical to that used by \cite{Pignatari2016}.

\begin{figure}
	\includegraphics[width=0.95\linewidth]{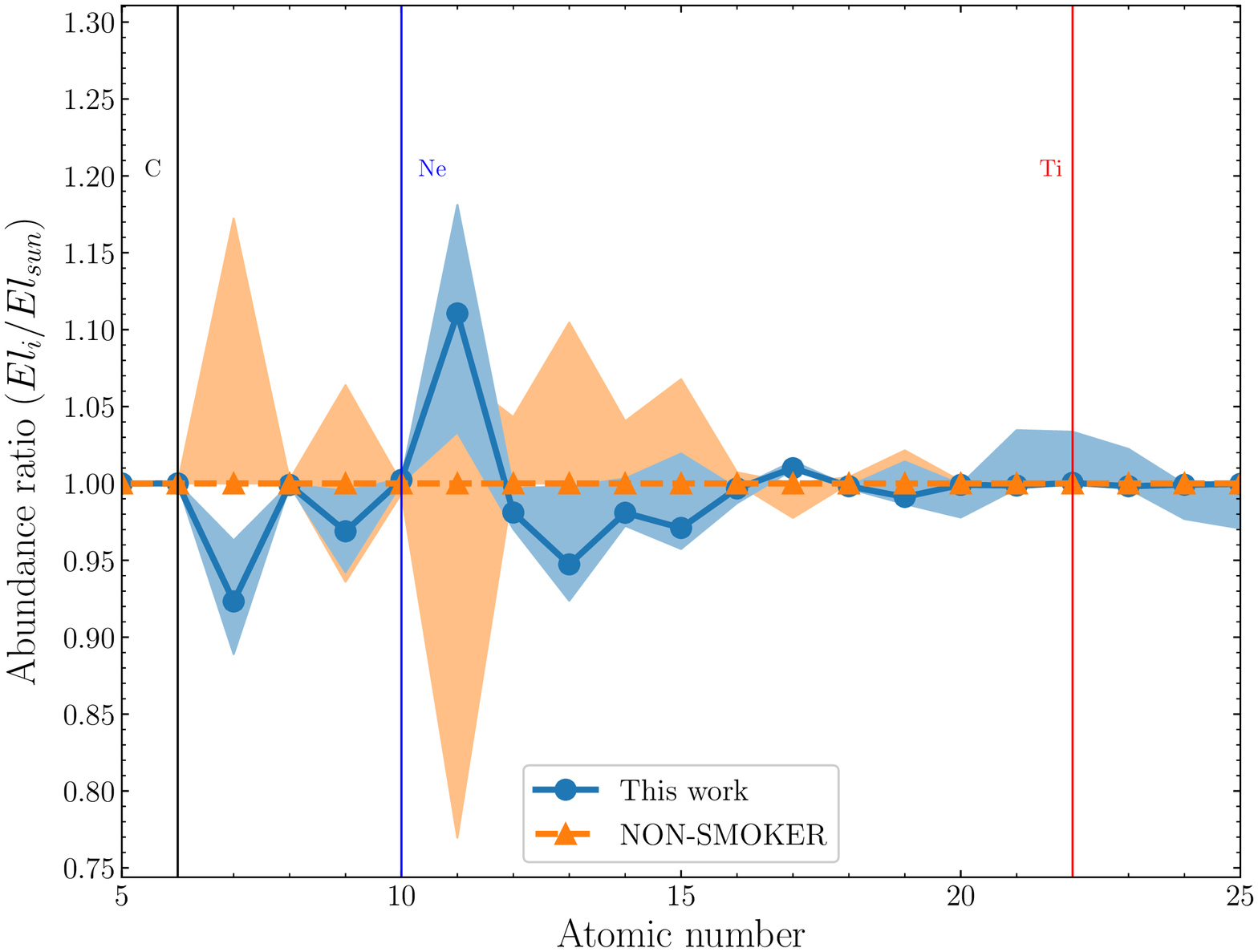}
	\caption{Elemental abundances determined using our new \NaMg{} rate (see fig. \ref{fig:CombinedReactionRate}) relative to the NON-SMOKER \citep{Rauscher2000} abundances for elements in the $Z = 5-25$ range, produced in the C-burning convective shell. The shaded regions correspond to uncertainty bands, assuming a best case factor of 2 for NON-SMOKER. The abundance of lighter isotopes is much better constrained than previously, most to less than 5\%.}
	\label{fig:60M_elemental_relative}
\end{figure}

For the impact of the new rate, \textsc{ppn} simulations are presented for eight rates: the original REACLIB rate, factors of 2 and 10 increase and decrease as upper and lower limits, the new experiment rate, and the upper and lower limits for this rate. The absolute differences of elemental abundances for elements with $Z = 5-25$ compared to REACLIB is small, as would be expected with the similar cross-sections, and so the data are plotted relative to REACLIB in figure \ref{fig:60M_elemental_relative}, with the shaded regions corresponding to the uncertainty of the rates. 

The uncertainty on the REACLIB rates is generally expected to be anywhere from a factor of 2 to 10. STARLIB, which provides uncertainties on all the rates, gives a factor of 10 uncertainty on the \NaMg rate \citep{Sallaska2013}. These uncertainties correspond to a change in $^{26}$Al of (+27\%,-14\%) for a factor of 2 and (+128\%,-29\%) for a factor of 10, and a change in $^{23}$Na of (+17\%,-23\%) for a factor of 2 and (+34\%,-70\%) for a factor of 10. In contrast, the uncertainties on the present experimental rate produce a change in both $^{26}$Al and $^{23}$Na of (+7\%,-7\%). In general the uncertainties on the abundances of the elements is constrained to within 5\% with the experimental rate, compared to approximately 20\% with the NON-SMOKER rates. The impact on intermediate-mass elements is negligible. However it should be noted that the largest contribution to the $^{26}$Al yield comes from explosive burning and the H-burning shells in massive stars, and so the contribution of the C convective shell to the total yield is negligible \citep{Limongi2006}

\begin{figure}
	\includegraphics[width=0.95\linewidth]{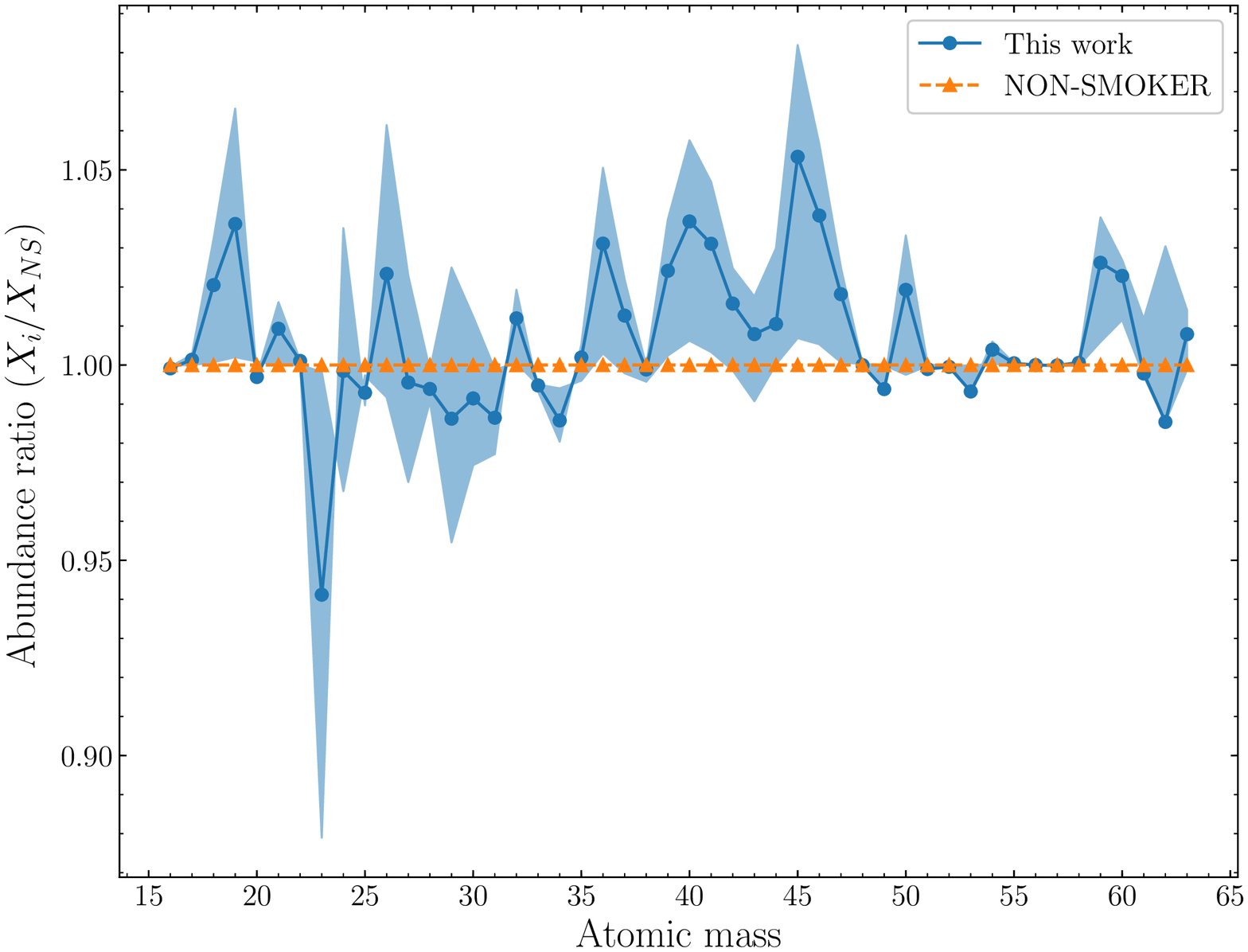}
	\caption{
		Ratios of abundances determined using our new \NaMg{} rate (see fig. \ref{fig:CombinedReactionRate})	to abundances determined using the NON-SMOKER rate \citep{Rauscher2000}, for the adopted type Ia supernova model (see text). The shaded region represents the impact on the predicted abundances of the uncertainty in our new rate.
	}
	\label{fig:IsoSN1a}
\end{figure}

The impact of our new \NaMg{} rate was also examined using a 2D Chandrasekhar-mass deflagration-to-detonation transition type Ia supernova explosion model \citep{Parikh2013}. As shown in fig \ref{fig:IsoSN1a}, the $^{23}$Na variation due to the uncertainty of our rate is within 15\%, and within 5\% for all other species.

\section{Summary}

A new combined experimental reaction rate for the \NaMg reaction has been calculated from three recent independent measurements \citep{AlmarazCalderon2015,Howard2015,Tomlinson2015}, and utilizing directly measured angular distributions for the observed cross-section to reduce systematic uncertainty. Impact on astrophysical scenarios has been evaluated in the carbon-burning convective-shell in massive stars,
and in type-Ia supernovae. The rate is found to be in good agreement with the statistical reaction rate previously used, but the uncertainty has been reduced from greater than a factor of 2 to 30\%. This uncertainty is dominated by uncertainties on the differential cross-section measurements themselves, and a conservative assumption on front-back asymmetry in the angular distributions which was not possible to observe. The abundance of $^{23}$Na is increased by a factor of 1.1, and $^{26}$Al is unchanged from HF predictions. Both abundances are constrained to within 7\%, this is compared to to an uncertainty of (+128\%,-29\%) for $^{23}$Na, and (+34\%,-70\%) for $^{26}$Al, using the factor of 10 uncertainty given by STARLIB \citep{Sallaska2013}. While the overall yield of $^{26}$Al has not been impacted by this work, the uncertainty on the \NaMg{} rate has been significantly reduced.

\acknowledgments
We acknowledge financial support from the European Research Council under ERC Starting grant LOBENA, No. 307447 and from the UK Science \& Technology Facilities Council under Grant No. ST/J000124/1.
This work benefited from support by the National Science Foundation under Grant No. PHY-1430152 (JINA Center for the Evolution of the Elements).
N. Hubbard would like to expect gratitude to the University of York High Performance Computing team for providing access to YARCC (the York Advanced Research Computing Cluster) on which the NuGrid calculations were performed.
M. Pignatari acknowledges significant support to NuGrid from the STFC (through the University of Hull's Consolidated Grant ST/R000840/1), and access to {\sc viper}, the University of Hull High Performance Computing Facility. M. Pignatari acknowledges the support from the ``Lendulet-2014'' Programme of the Hungarian Academy of Sciences (Hungary) and the ERC Consolidator Grant (Hungary) funding scheme (Project RADIOSTAR, G.A. n. 424560).
This article is based upon work from the ChETEC COST Action (CA16117), supported by COST (European Cooperation in Science and Technology), and the UK network BRIDGCE.

\software{ppn, mppnp, nugridpy \citep{NuGridPaper}, CINA \citep{CINA}, exp2rate \citep{exp2rate}, DDT Model \citep{Parikh2013}}

\bibliographystyle{aasjournal}
\bibliography{library}

\end{document}